%% file: main.tex
\documentclass[journal]{IEEEtran}
\usepackage[ruled, linesnumbered]{algorithm2e}
\usepackage{lineno,hyperref}
\usepackage{newtxtext,newtxmath}
\usepackage{amsmath}
\usepackage{multirow}
\modulolinenumbers[5]
\usepackage{caption}
\usepackage{hyperref}
\usepackage{threeparttable}
\usepackage[font=small,labelfont=bf]{caption}

\newcommand\eat[1]{}

\usepackage{todonotes}
\usepackage{soul}
\usepackage{amsmath}

\usepackage{tikz}
\usepackage{lipsum}

\usepackage{multirow}
\usepackage{cellspace}
\usepackage{hhline}
\usepackage{stackengine}
\setstackEOL{\\}

\makeatletter
\def\ps@IEEEtitlepagestyle{%
  \def\@oddfoot{\mycopyrightnotice}%
  \def\@evenfoot{}%
}
\def\mycopyrightnotice{%
  {\footnotesize  This paper has been accepted for publication by the IEEE Transactions on Multimedia. The copyright is with the IEEE. DOI: 10.1109/TMM.2020.3006415\hfill}
  \gdef\mycopyrightnotice{}
}

\ifCLASSINFOpdf
\else
\fi
\begin{document}
%
\title{An Automated and Robust Image Watermarking Scheme Based on Deep Neural Networks}
%
%
%

\author{Xin Zhong, 
        Pei-Chi Huang, 
        Spyridon Mastorakis, 
        Frank Y. Shih,~\IEEEmembership{Senior Member,~IEEE}
\thanks{Xin Zhong, Pei-Chi Huang and Spyridon Mastorakis are with Department of Computer Science, University of Nebraska Omaha, Omaha, NE, 68182 USA e-mail:\tt\small~\href{\{xzhong, phuang, smastorakis\}@unomaha.edu}{\{xzhong, phuang, smastorakis\}@unomaha.edu} }
\thanks{Frank Y. Shih is with Department of Computer Science, New Jersey Institute of Technology, Newark, NJ 07102 USA e-mail:\tt\small~\href{shih@njit.edu}{shih@njit.edu}}
}

\maketitle

\newcommand{\spyros}[1]{\textcolor{blue}{#1}}

\maketitle

\begin{abstract}
\input{Section/abstract.tex}
\end{abstract}


%
\IEEEpeerreviewmaketitle

\section{Introduction}
\input{Section/intro.tex}

\section{Related Work}
\label{sec:related}
\input{Section/related_work.tex}

\section{Proposed Image Watermarking Scheme}
\label{sec:tech_details}
\input{Section/tech_detail.tex}

\section{Experiments and Analysis}
\label{sec:eval}
\input{Section/experiment.tex}

\section{Applications of the Proposed Watermarking Scheme}
\label{sec:apps}
\input{Section/app.tex}

\section{Conclusion}
\label{sec:conclusion}
\input{Section/conclusion.tex}

\small
\bibliographystyle{IEEEtran}
\bibliography{ref.bib}


%




\ifCLASSOPTIONcaptionsoff
  \newpage
\fi

\end{document}

%% file: Section/abstract.tex
Digital image watermarking is the process of embedding and extracting a watermark covertly on a cover-image. To dynamically adapt image watermarking algorithms, deep learning--based image watermarking schemes have attracted increased attention during recent years. However, existing deep learning--based watermarking methods neither fully apply the fitting ability to learn and automate the embedding and extracting algorithms, nor achieve the properties of robustness and blindness simultaneously. In this paper, a robust and blind image watermarking scheme based on deep learning neural networks is proposed. To minimize the requirement of domain knowledge, the fitting ability of deep neural networks is exploited to learn and generalize an automated image watermarking algorithm. A deep learning architecture is specially designed for image watermarking tasks, which will be trained in an unsupervised manner to avoid human intervention and annotation. To facilitate flexible applications, the robustness of the proposed scheme is achieved without requiring any prior knowledge or adversarial examples of possible attacks. A challenging case of watermark extraction from phone camera--captured images demonstrates the robustness and practicality of the proposal. The experiments, evaluation, and application cases confirm the superiority of the proposed scheme.

%% file: Section/intro.tex
Digital image watermarking refers to the process of embedding and extracting information covertly on a cover-image. The data (\textit{i.e.,} the watermark) is hidden into a cover-image to create a to-be-transmitted marked-image. The marked-image does not visually reveal the watermark, and only the authorized recipients can extract the watermark information correctly. The techniques of image watermarking can be applied for various applications. Based on different target scenarios, the watermark information can be presented in different forms; for example, the watermark can be some random bits or electronic signatures for image protection and authentication~\cite{berghel1996protecting}, or some hidden messages for covert communication~\cite{cox1999watermarking}. In addition, the watermark can be encoded for different purposes, such as added security with encryption methods or restoring information integrity with error correction codes during a cyberattack~\cite{shih2017digital}.

For copyright protection, classic image watermarking research~\cite{cox1997secure} only focuses on single-bit extractions, where the output indicates whether an image contains a watermark or not. To enable a wide range of applications, modern image watermarking research primarily focuses on multi-bit scenarios that extract the entire communicative watermark information~\cite{shih2017digital, cox2007digital}. 
Typically, many factors should be considered in an image watermarking scheme, such as the fidelity of the marked-image and the watermark’s undetectability to computer analysis. The proposed image watermarking scheme not only satisfies these factors but achieves the robustness of its priority: the watermark should survive even if the marked-image is degraded or distorted. Ideally, a robust image watermarking scheme keeps the watermark intact under a designated class of distortion without any assistance of techniques. However, in practice, the watermark is extracted approximately in many attacking scenarios, and various encoding methods can be applied for its restoration~\cite{kang2003dwt}. Achieving robustness is a major challenge in a blind image watermarking scheme, where the extraction must be performed without any information about the original cover-image.

Due to the limited scope of manual design, the traditional image watermarking methods encounter difficulties; for example, extraction can only tolerate certain types of distortions in robust watermarking schemes, or the watermark itself can only resist a limited range of computer analysis in undetectable watermarking schemes~\cite{craver1998resolving}. To break through these drawbacks, incorporating deep learning into image watermarking has attracted increased attention in recent years~\cite{kandi2017exploring}. Deep learning, as a representation learning method, has enabled significant improvements in computer vision through its ability to fit and generalize complex features. The major advantage of deep learning methodologies for image watermarking is that they perform image watermarking in a more adaptive manner by dynamically learning algorithms to extract both high- and low-level features for image watermarking from multiple instance big data.

Recent research on image watermarking tasks with deep neural networks has emerged~\cite{vukotic2018deep,li2019novel,fierro2019robust,papernot2016limitations}, but there still exist challenging issues. For example, it is difficult to fully utilize the fitting ability of deep neural networks to automatically learn and generalize both the watermark embedding and extracting processes. Also, labeling the ground truth for an image watermarking task can be ill-defined or time-consuming. Finally, achieving robustness and blindness simultaneously without prior knowledge of adversarial examples~\cite{papernot2016limitations} remains unexplored. 

To address the above challenges, we present an automated, blind, and robust image watermarking scheme based on deep learning neural networks. The contribution of this paper is threefold. First, the fitting ability of deep neural networks is exploited to automatically learn image watermarking algorithms, facilitating an automated system without requiring domain knowledge. Second, the proposed deep learning architecture can be trained in an unsupervised manner to reduce human intervention, which is suitable for image watermarking. Finally, experimental results demonstrate the robustness and accuracy of the proposed scheme without using any prior knowledge or adversarial examples of possible attacks. 

The remainder of this paper is organized as follows. The related work is described in Section~\ref{sec:related}. The proposed scheme is presented in Section~\ref{sec:tech_details}. Experiments and analysis are presented in Section~\ref{sec:eval}. Applications of the proposed watermarking scheme are discussed in Section~\ref{sec:apps}. Finally, conclusions are drawn in Section~\ref{sec:conclusion}.

\vspace{-0.3cm}

%% file: Section/related_work.tex
This section provides a detailed analysis of recent reports. Table~\ref{Table: analytic_comparison} shows the analytical comparison of our proposed scheme with the start-of-the-art deep learning--based image watermarking schemes.

In handcrafted watermarking algorithms, various optimization methods have been applied to adapt the embedding parameters, and this research direction has attracted attentions in recent years~\cite{huang2019enhancing,su2018snr,chen2018novel}. Consequently, exploring the optimization ability of deep learning models for adaptive and automated image watermarking is of great interest.
However, compared to significant advancements on image steganography with deep neural networks~\cite{wang2018sstegan,baluja2017hiding}, deep learning--based image watermarking is still in its infancy.

Kandi \textit{et al.}~\cite{kandi2017exploring} applied two convolutional autoencoders to reconstruct a cover-image. In a marked-image, the pixels produced by the first autoencoder indicate bits with the value of zero, and the pixels produced by the second autoencoder indicate bits with the value of one, hence developing a non-blind binary watermarking scheme. Vukotic \textit{et al.}~\cite{vukotic2018deep} developed a deep learning--based, single-bit watermarking scheme by embedding through designed adversarial images and extracting via the first layer of a trained deep learning model. Li \textit{et al.}~\cite{li2019novel} embedded a watermark into the discrete cosine domain by traditional algorithms~\cite{cox1997secure} and applied convolutional neural networks to facilitate the extraction. 

Besides the single-bit and multi-bit watermarking, attempts were also reported in special scenarios. For example, for zero-watermarking, where a master share is sent separately from the image, Fierro-Radilla \textit{et al.}~\cite{fierro2019robust} applied convolutional neural networks to extract the required feature from the cover-image and linked these features with the watermark to create a master share. For the scenario of template-based watermarking, Kim \textit{et al.}~\cite{kim2018convolutional} embedded a handcrafted template by using a classic additive method and estimated possible distortion parameters by comparing the extracted template to the original one with the help of convolutional neural networks. Thus far, the existing deep learning--based image watermarking schemes do not fully apply the fitting ability of deep neural networks to learn and generalize the embedding and extracting algorithms. 

Furthermore, due to the fragility of deep neural networks~\cite{papernot2016limitations}, a modified image as an input to a trained deep learning model may cause failure. In other words, robustness is a major challenge in deep--learning based image watermarking because noise or modifications of the marked-image can result in extraction failures. Mun \textit{et al.}~\cite{mun2019finding} proposed to solve this issue by proactively including noisy marked-images as adversarial examples in the training phase. However, enumerating all types of attacks and their combinations may not be practically feasible.

To the best of our knowledge, our proposed scheme is the first method that explores the ability of deep neural networks in automatically learning and generalizing both watermark embedding and extracting algorithms, while achieving robustness and blindness simultaneously. 

\begin{table*}
\centering
\caption{An analytical comparison between the proposed scheme and state-of-the-art image watermarking methods applying deep neural networks.}
\label{Table: analytic_comparison}
\begin{tabular}{ l || c | c | l | c}
\hline \hline
\textbf{Method}                                             & \textbf{\begin{tabular}[c]{@{}c@{}}Learning \\ Watermarking \\ Algorithms?\end{tabular}} & \textbf{Blind} & \multicolumn{1}{c|}{\textbf{Robust}} & 
\textbf{\begin{tabular}[c]{@{}c@{}}Extraction \\ Type\end{tabular}}
\\  \hline \hline
Kandi \textit{et al.}~\cite{kandi2017exploring}  & No & No  & Robust to common image processing attacks  & Multi-bit    \\ \hline
Vukotic \textit{et al.}~\cite{vukotic2018deep}                                                    & Learning extraction & Yes            &  Robust to Rotation, JPEG, and Cropping     & Single-bit               \\ \hline
Li \textit{et al.}~\cite{li2019novel}  & No  & No  & No   & Multi-bit                \\ \hline
Fierro-Radilla \textit{et al.}~\cite{fierro2019robust} & No    & No  & Robust to common image processing attacks & Zero-watermarking \\ \hline
Kim \textit{et al.}~\cite{kim2018convolutional}  & Assisting extraction & No   & Focus on geometric attacks  &    Template-based watermarking            \\ \hline
Mun \textit{et al.}~\cite{mun2019finding}  & Learning extraction & Yes    & Robust to all enumerated attacks during training & Multi-bit                \\ \hline
Ours                                                             & Yes                                                                                      & Yes            & Robust to common image processing attacks        & Multi-bit                \\ \hline \hline
\end{tabular}
\end{table*}

%% file: Section/tech_detail.tex
We revisit the typical design of an image watermarking scheme and present the overview architecture of our scheme in Section~\ref{subsec:typical_design}. Then, we present the loss function design and the scheme objective in Section~\ref{subsec:objective}. Finally, the detailed structure of the proposed model is described in Section~\ref{subsec:structure_detail}.

\subsection{The Overview Architecture of the Proposed Scheme}\label{subsec:typical_design}

\begin{figure*}
\centering
    \includegraphics[width=0.7\linewidth]{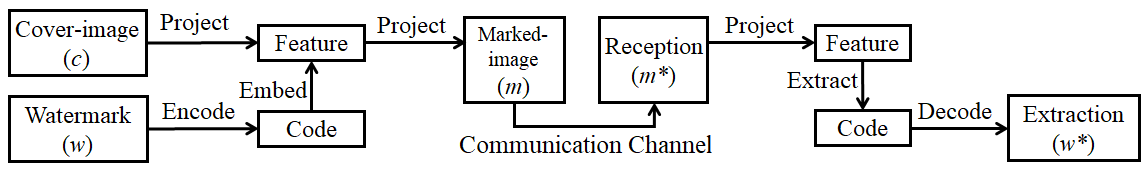}
    \centering
    \caption{The traditional design of an image watermarking scheme.}
    \vspace{-0.5cm}
    \label{fig:wm_general}
\end{figure*}

The traditional design of an image watermarking scheme is shown in~Fig.\ref{fig:wm_general}. A watermark (denoted as $w$) is embedded into a cover-image (denoted as $c$) to produce a marked-image (denoted as $m$) that looks similar to $c$ and is transported through a communication channel. Then, the receiver extracts the watermark data (denoted as $w^*$) from the received marked-image (denoted as $m^*$) that could be a modified version of $m$ if some distortions or attacks are occurred during the transmission.

To embed $w$ into $c$, typically, the first step is to project $c$ into one of its feature spaces in spatial, frequency, or other domains. Next, $w$ is encoded and embedded into the feature space of $c$. The embedded feature space is projected back into the cover-image space to create a marked-image $m$. Inversely, the watermark extraction is to project the marked-image reception $m^*$ to the same feature space and then extract and decode the watermark information. Based on different target applications, traditional image watermarking methods manually design the projection, embedding, extraction, encoding, and decoding functions. As the criteria of a design, an image watermarking scheme often highlights its fidelity (\textit{i.e.,} high similarity between the $m$ and $c$) and robustness (\textit{i.e.,} keeping the integrity of $w^*$ when $m^*$ is distorted).

\begin{figure*}[h!]
    \centering
    \includegraphics[width=0.7\linewidth]{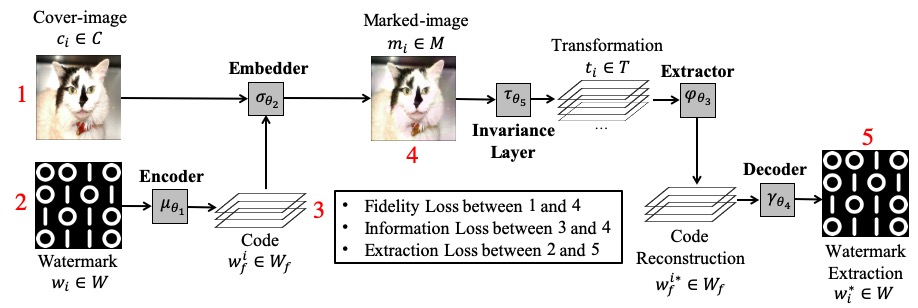}
    \caption{The overall architecture of the proposed image watermarking scheme.}
    \vspace{-0.3cm}
    \label{fig:overall_architecture}
\end{figure*}

Traditional image watermarking methods perform competently through hand-designed algorithms; however, it remains challenging to automatically learn these algorithms without complete dependence upon careful design. To tackle this difficulty, we propose a novel scheme which develops a deep learning model to automatically learn and generalize the embedding, the extraction, and the invariance functions in image watermarking. 
Fig. \ref{fig:overall_architecture} illustrates the overall architecture of the proposed scheme with some example images. Given two input spaces that are all the possible inputs of watermark images and cover-images ($W$ and $C$, respectively), neural network $\mu_{\theta_1}$ parameterized by $\theta_1$ is applied to learn a function that encodes $W$. $W_f$, the encoded space of $W$, not only enlarges $W$ to prepare for the next-step concatenation, but also brings some redundancy, decomposition, and perceivable randomness to help information protection and robustness. Like the embedding process in traditional watermarking where an encoded $w$ is inserted into a feature space of $c$, in the proposed scheme, an embedder that takes $W_f$, $C$ as inputs and produces the marked-image is fit by the neural network $\sigma_{\theta_2}$ parameterized by $\theta_2$.
The marked-image space is named as $M$. To handle possible distortions, a neural network $\tau_{\theta_5}$ parameterized by $\theta_5$ is introduced to learn to convert $M$ to its enlarged and redundant transformed-space $T$. After the transformation, $T$ preserves information about $W_f$ and rejects other irrelevant information, such as noises on $M$, therefore providing robustness.
Finally, the inverse watermark reconstruction functions are fitted by two neural network components, $\varphi_{\theta_3}$ and $\gamma_{\theta_4}$ with trainable parameters $\theta_3$ and $\theta_4$, that extract $W_f$ from $T$ and decode $W$ from $W_f$, respectively.

To compare the proposed scheme in Fig.~\ref{fig:overall_architecture} with the traditional design in Fig.\ref{fig:wm_general}, one can observe that the neural network components fit and optimize the image watermarking process dynamically. Encoder and decoder networks (denoted as $\mu_{\theta_1}$ and $\gamma_{\theta_4}$, respectively) fit the watermark encoding and decoding functions. An embedder (denoted as $\sigma_{\theta_2}$) projects $C$ into a feature space, embeds $W_f$ into the space, and projects to the marked-image space $M$. An extractor (denoted as $\varphi_{\theta_3}$) inverses all the processes in $\sigma_{\theta_2}$, and $\tau_{\theta_5}$ handles the distortion during the transmission through a communication channel. More details of the architectures are described in Section~\ref{subsec:structure_detail}.

Compared to an autoencoder~\cite{hinton2006reducing}, where an input space is transformed to a representative and intermediate feature space and the original input is recovered from this feature space, the proposed scheme takes two spaces $C$ and $W$ as inputs, produces an intermediate marked-image space $M$, and recovers $W$ from $M$.
The recovery ability of autoencoders, \textit{i.e.,} an exact reconstruction of the input with appropriate features extracted by the deep neural networks, secures the feasibility of watermark extraction in the proposed scheme. The reconstruction requires only $M$ without the need of $W$ and $C$, enabling the {\em blindness} of the proposed scheme. A feature space in an autoencoder is often learned through a bottleneck for the dimensionality compression, but the proposed scheme learns equal-sized or over-complete representations to achieve high {\em robustness} and {\em accuracy} of watermark extraction. 

\subsection{The Loss Function and Scheme Objective}\label{subsec:objective}
The entire architecture is trained as a single deep neural network with several loss terms designed for image watermarking. Given the data samples $w_i \in W$, $i=1,2,3,\ldots$ and $c_i \in C$, $i=1,2,3,\ldots$, the proposed scheme can be trained in an unsupervised manner. There are two inputs $w_i$ and $c_i$, and two outputs $w_i^*$ and $m_i$ in the proposed deep neural network. For the output $w_i^*$, an extraction loss that minimizes the difference between $w_i^*$ and $w_i$ is computed to ensure full extraction of the watermark. For the output $m_i$, a fidelity loss that minimizes the difference between $m_i$ and $c_i$ is computed to enable watermarking invisibility. For the output $m_i$, we also compute an information loss that forces $m_i$ to contain the information of $w_i$. To achieve this, we maximize the correlation between feature maps of $w_f^i$ and feature maps of $m_i$, where the feature maps are the outputs of convolutional layers in the proposed architecture, and the feature maps of $w_f^i$ and $m_i$, \textit{i.e.}, $B_1$ and $B_2$, are illustrated in Figs.~\ref{fig:parameters} and~\ref{fig:conv_block}. Denoting the parameters to be learned as $\vartheta=[\theta_1, \theta_2, \theta_3, \theta_4, \theta_5]$, the loss function $L(\vartheta)$ of the proposed scheme can be expressed as:
\begin{equation}\label{equ:lossfunc}
    L(\vartheta) = \lambda_1 \| w_i^*-w_i \|_1 + \lambda_2 \|m_i-c_i \|_1  + \lambda_3 \psi(m_i,w_f^i), 
\end{equation}
where $\lambda_i$, $i=1, 2, 3$ is the weight factor and $\psi$ is a function computing the correlation given as:
\begin{equation}\label{equ:loss_correlation}
\begin{aligned}
\psi(m_i,w_f^i ) = & \frac{1}{2} (\|g(B_1(w_f^i)), g(B_1(m_i)) \|_1 +  \\
                   &  \|g(B_2 (w_f^i)), g(B_2 (m_i)) \|_1),
\end{aligned}
\end{equation}
where $g$ denotes the Gram matrix that contains all possible inner products. By minimizing the distance between the Gram matrices of the feature maps of $m_i$ and $w_f^i$ produced by intermediate layer outputs $B_1$ and $B_2$, we maximize their correlation. As the feature producers, the annotation of $B_1$ and $B_2$ is presented in Fig.~\ref{fig:conv_block}, and the convolution block $B$ that contains $B_1$ and $B_2$ is annotated in Fig.~\ref{fig:parameters}. More discussions will be presented in Section~\ref{subsec:structure_detail}.

In Eq.~\ref{equ:lossfunc}, each two of the fidelity loss, information loss, and extraction loss terms can be a trade-off for image watermarking. For example, minimizing the fidelity loss term to zero means that $m_i$ is identical to $c_i$. However, there is no embedded information in $m_i$ in this case, so the extraction of $w_i$ will fail. To allow some imperfectness of the loss terms, the mean absolute error (\textit{i.e.,} the L1 norm) is selected to highlight the overall performance rather than a few outliers.

With regularization, the proposed scheme objective is represented as $L(\vartheta)+\lambda_4 P$, where $P$ is the penalty term to achieve robustness as in Eq.~\ref{equ:reg_term}, and $\lambda_4$ is the weight controlling the strength of the regularization term. The deep neural network needs to learn the parameter $\vartheta^*$ that minimizes $L(\vartheta)+\lambda_4 P$:
\begin{equation}\label{equ:objective}
   \vartheta^* = argmin_{\vartheta} [L(\vartheta)+ \lambda_4 P].         
\end{equation}
In the backpropagation during training, the term $\lambda_1 \| w_i^*-w_i \|_1$ is applied by all the components of the proposed architecture in their weight updates, while only $\mu_{\theta_1}$ and $\sigma_{\theta_2}$ apply terms $\lambda_2 \| m_i-c_i \|_1$ and $\lambda_3 \psi(m_i,w_f^i )$ to their weight updates. This enables $\mu_{\theta_1}$ and $\sigma_{\theta_2}$ to encode and embed the information in a way that $\varphi_{\theta_3}$ and $\gamma_{\theta_4}$ are able to extract and decode the watermark.

\subsection{Detailed Structure of the Proposed Neural Networks Model}
\label{subsec:structure_detail}
\begin{figure}
    \centering
    \includegraphics[width=1\linewidth]{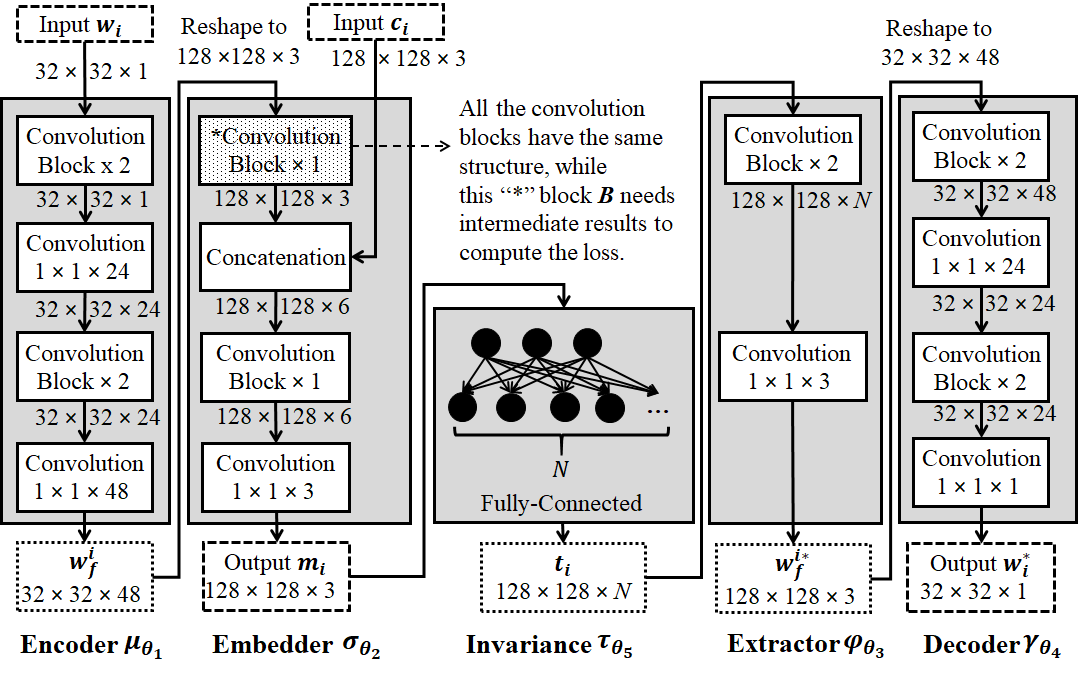}
    \vspace{-0.5cm}
    \caption{The detailed components of the proposed watermarking scheme: the Encoder $\mu_{\theta_1}$, the Embedder $\sigma_{\theta_2}$, the Invariance Layer $\tau_{\theta_5}$, the Extractor $\varphi_{\theta_3}$, and the Decoder $\gamma_{\theta_4}$. Every structure of the convolution block is the same, but only block marked with ''*'' needs intermediate results to compute the loss.} 
    \vspace{-0.7cm}
    \label{fig:parameters}
\end{figure}
This subsection describes the major components design of neural networks: $\mu_{\theta_1}$, $\sigma_{\theta_2}$, $\varphi_{\theta_3}$, $\gamma_{\theta_4}$ and $\tau_{\theta_5}$ in more detail. The overall design is modularized and illustrated in Fig.~\ref{fig:parameters}. If we single out two pairs ($\mu_{\theta_1}$, $\gamma_{\theta_4}$) and ($\sigma_{\theta_2}$, $\varphi_{\theta_3}$), we can find that each pair is conceptually symmetrical. The watermark is considered as a $32 \times 32 \times 1$ binary image, and the cover-image is a $128 \times 128 \times 3$ color image in this description. One might adapt and customize the image sizes based on different target applications. 

\subsubsection{The Encoder $\mu_{\theta_1}$ and the Decoder $\gamma_{\theta_4}$}
Given the samples $w_i,$ $i=1,2,3,...$ from the input space $W$, the encoder $\mu_{\theta_1}$ learns a function that encodes $W$ to its code $W_f$. Inversely, the decoder $\gamma_{\theta_4}$ learns the decoding function from $W_f$ to $W$ with samples $w_f^{i*}$, $i=1,2,3,...$. The encoder $\mu_{\theta_1}$ successively increases a $32 \times 32 \times 1$ binary watermark image to $32 \times 32 \times 24$ and $32 \times 32 \times 48$, and the decoder $\gamma_{\theta_4}$ successively decreases the $32 \times 32 \times 48$ feature space back to a $32 \times 32 \times 1$ binary watermark image. The reason to train this channel-wise increment is two-fold. First, it produces a $128 \times 128 \times 3$ $w_f^i$ that has the same width and height as the cover-image, so that we can concatenate a feature map of $w_f^i$ and $c_i$ along their channel dimension. Each of $w_f^i$ and $c_i$ will contribute equally to the $128 \times 128 \times 6$ concatenated matrix used in the embedder $\sigma_{\theta_2}$. Thus, we are evenly weighing the watermark and the cover-image. 
Second, this $32 \times 32 \times 1$ to $32 \times 32 \times 48$ increment introduces some redundancy, decomposition, and perceivable randomness to $W$, which helps information protection and robustness.

\subsubsection{The Embedder $\sigma_{\theta_2}$ and the Extractor $\varphi_{\theta_3}$} 

The embedder $\sigma_{\theta_2}$ applies the convolution block $B$ to extract a $128 \times 128 \times 3$ to-be-embedded feature map of $w_f^i$ that is concatenated along the channel dimension with the cover-image. Directly applying $c_i$, while only applying a feature map of $w_f^i$, helps $c_i$ to dominate the appearance. The $128 \times 128 \times 6$ concatenation is fed into another convolution block to produce $m_i$. The extractor $\varphi_{\theta_3}$ inverses the process by two successive convolution blocks.

\begin{figure}[h!]
    \centering
    \includegraphics[width=1\linewidth]{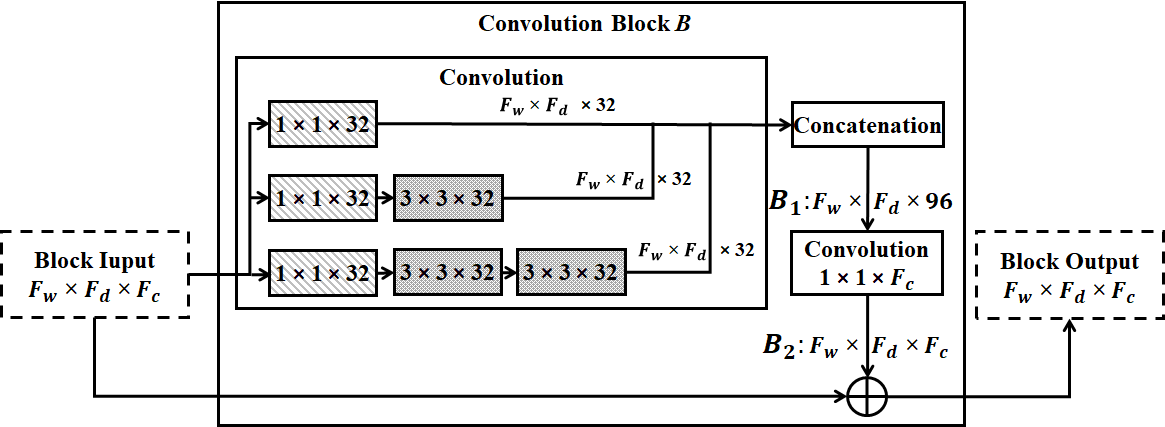}
    \caption{The architecture design of a convolution block. The block input/output size is denoted as the size of height ($F_w$), the width ($F_d$), and the channel ($F_c$), respectively.}
    \label{fig:conv_block}
\end{figure}

To capture various scales of features for image watermarking, the inception residual block~\cite{szegedy2017inception} is applied. It consists of a $1 \times 1$, a $3 \times 3$, and a $5 \times 5$ convolution, as well as a residual connection that sums up the features and the input itself. In the proposed structure, each convolution has 32 filters, and the 5 × 5 convolution is replaced by two 3 × 3 convolutions for efficiency. The 32-channel feature maps produced by different convolution paths are concatenated along the channel dimension to form a 96-channel feature, and a 1 × 1 convolution is applied to convert the 96-channel feature back to the original input channel size for the summation in the residual connection. The architecture of a convolution block is shown in Fig.~\ref{fig:conv_block}, where $F_w$, $F_d$, and $F_c$, respectively, denote the size of height, width, and channel.

All the convolution blocks in Fig.~\ref{fig:parameters} have the same inception residual structure as shown in Fig.~\ref{fig:conv_block}. In the case of the "*" convolution block $B$ of Fig.~\ref{fig:parameters}, the annotated intermediate results $B_1$ and $B_2$ of Fig.~\ref{fig:conv_block} are applied in Eq.~\ref{equ:loss_correlation}. Specifically, block $B$ extracts features not only from its input $w_f^i$ in the architecture, but also from $m_i$. The annotated $F_w \times F_d \times 96$ and $F_w \times F_d \times F_c$ feature maps are the intermediate results $B_1$ and $B_2$, respectively. 

\subsubsection{The Invariance Layer $\tau_{\theta_5}$}
The invariance layer is the key component to provide robustness in the proposed image watermarking scheme. Using a fully-connected layer, $\tau_{\theta_5}$ learns a transformation from space $M$ to an over-complete space $T$, where the neurons are activated sparsely. The idea is to redundantly project the most important information from $M$ into $T$ and to deactivate the neural connections of the areas on $M$ irrelevant of the watermark, thus preserving the watermark even if there is noise or distortion that modified a part of $M$. As shown in Fig.~\ref{fig:parameters}, $\tau_{\theta_5}$ converts a $3$-color-channel instance $m_i$ of $M$ into an $N$-channel ($N \geq 3$ for non-compression) instance $t_i$ of $T$, where $N$ is the redundant parameter. Increasing $N$ results in increased redundancy and decomposition in $T$, which provides higher tolerance of the errors in $M$ and enhances robustness. 

Based on the contractive autoencoder~\cite{rifai2011contractive}, $\tau_{\theta_5}$ employs a regularization term that is obtained by the Frobenius norm of the Jacobian matrix of a layer's outputs with regards to its inputs. Mathematically, the regularization term $P$ is given as:
\begin{equation}
P = \Sigma_{i,j} \left(\frac{\partial h_j(X)}{\partial X_i} \right)^2, 
\end{equation}
where $X_i$ denotes the $i$-th input and $h_j$ denotes the output of the $j$-th hidden unit of the fully connected layer. Similar to a common gradient computation, the Jacobian matrix can be written as:
\begin{equation}
\frac{\partial h_j(X)}{\partial X_i} = \frac{\partial A(\omega_{ji} X_i)}{\partial \omega_{ji} X_i} \omega_{ji},
\end{equation}
where $A$ is an activation function and $\omega_{ji}$ is the weight between $h_j$ and $X_i$. We set $A$ as the hyperbolic tangent ($\tanh$) for strong gradients and bias avoidance~\cite{lecun2012efficient}, and hence $P$ can be computed as:
\begin{equation}\label{equ:reg_term}
P = \Sigma_j (1-h_j^2)^2 \Sigma_i (\omega_{ji}^T)^2.
\end{equation}

If the value of $P$ is minimized to zero, all weights $\omega$ in $\tau_{\theta_5}$ will be zero, so that the output of $\tau_{\theta_5}$ will be always zero no matter how we change the inputs $X$. Thus, minimizing $P$ alone will cause the rejection to all the information from the inputs $m_i$. Therefore, we place $P$ as a regularization term in the total loss function to preserve useful information related to the loss terms of image watermarking, while rejecting all other noise and irrelevant information. In this way, we achieve robustness without prior knowledge of possible distortion.

Remarkably, each color channel in $m_i$ is treated as a single input unit to significantly improve the computational efficiency. For example, if we treat one pixel as an input, a $128 \times 128 \times 3$ marked-image will have 49,152 input units. Setting the redundant parameter \textit{N} to its smallest value 3 will imply $49,152 \times 3$ (= 147,456) units in the fully-connected invariance layer $\tau_{\theta_5}$, which requires at least $49,152 \times 147,456$ (= 7,247,757,312) parameters. This significantly lowers the efficiency. On the other hand, treating one color channel as an input unit considers only 3 input units for an RGB marked-image, which enables faster computation with much fewer parameters as well as a much larger \textit{N} to enable higher redundancy for higher robustness.

%% file: Section/experiment.tex
This section experimentally analyzes quantitative and analytical evaluation of the proposed deep learning--based image watermarking scheme. Section~\ref{subsec:dataset} introduces our data preparation, and Section~\ref{subsec:train_test} presents the experimental design, training and validation. To validate our proposed image watermarking approach, Section~\ref{subsec:synthetic_images} provides special testing experiments on synthetic images, and Section~\ref{subsec:robust} shows the robustness in different distortion. A feasibility case study on the scenario of watermark extraction from phone camera pictures is also presented in Section~\ref{subsec:feasibility_test}.

\subsection{Preparation of Datasets}\label{subsec:dataset}
The proposed deep learning--based image watermarking architecture was trained as a single deep neural network. ImageNet~\cite{russakovsky2015imagenet} was rescaled to size $128 \times 128$ with RGB channels and then used as the cover-images. The binary version of CIFAR \cite{krizhevsky2009learning} with its original size $32 \times 32$ was used as the watermarks because the proposed architecture used $32 \times 32$ binary watermark images. Both datasets contain more than millions of images such that the proposed scheme during training can be introduced by a large scope of instances. For a validation set after each training epoch, $10,000$ images from each dataset that were not used during the training phase are separated. 

The testing is performed on 10,000 images (rescaled to $128 \times 128$) from the Microsoft COCO dataset \cite{lin2014microsoft} as the cover-images, and 10,000 images of the testing set of the binary CIFAR as the watermarks. Both the testing cover-images and watermarks were not applied in the training to demonstrate that the proposed scheme learns and generalizes the watermarking algorithms without over-fitting to the training samples.

\subsection{Training, Validation and Testing of the Proposed Model}\label{subsec:train_test}
\begin{figure}[h!]
    \vspace{-0.4cm}
    \centering
    \includegraphics[width=1\linewidth]{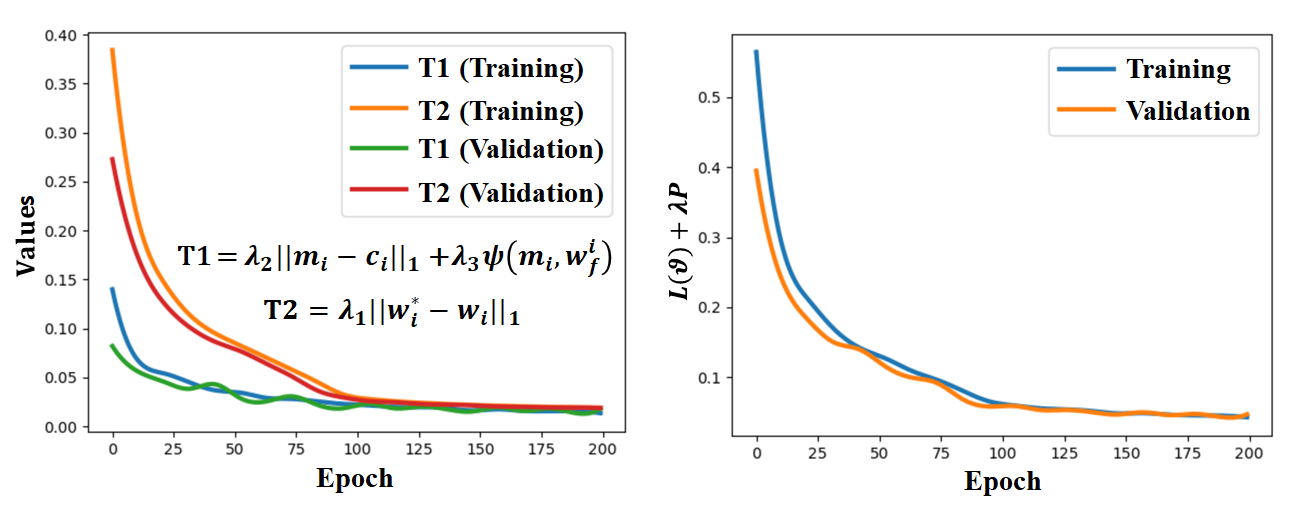}
    \caption{The loss values during training and validation.}
    \vspace{-0.2cm}
    \label{fig:training}
\end{figure}

As described in Section~\ref{sec:tech_details}, the proposed image watermarking scheme is trained as a single and deep neural network. The ADAM optimizer~\cite{kingma2014adam}, which adopts a moving window in the gradient computation, is applied, for its ability of continuous learning after large epochs. The training and validation of the proposed scheme are shown in 
Fig. \ref{fig:training}, where the values of the terms in the loss (Eq. \ref{equ:lossfunc}) and objective (Eq. \ref{equ:objective}) during $200$ epochs are presented. During both training and validation, the terms T1 and T2 (defined in Fig. \ref{fig:training}) in $L(\vartheta)$ converge smoothly below 0.015, and $L(\vartheta)+\lambda_4 P$ converges below 0.03, indicating a proper fit. Term T1 has slightly more errors because when carrying the watermark, a marked-image cannot be completely identical to a cover-image. $\lambda_1$, $\lambda_2$, and $\lambda_3$ are all set to be 1 to equally weigh the terms, and $\lambda_4$ is set to be 0.01 as suggested in \cite{rifai2011contractive}. All the layers apply the rectified linear unit (ReLU) as the activation function apart from the outputs (marked-image and watermark extraction), which use sigmoid to limit the range to (0, 1).

\begin{figure}
    \centering
    \includegraphics[width=0.9\linewidth]{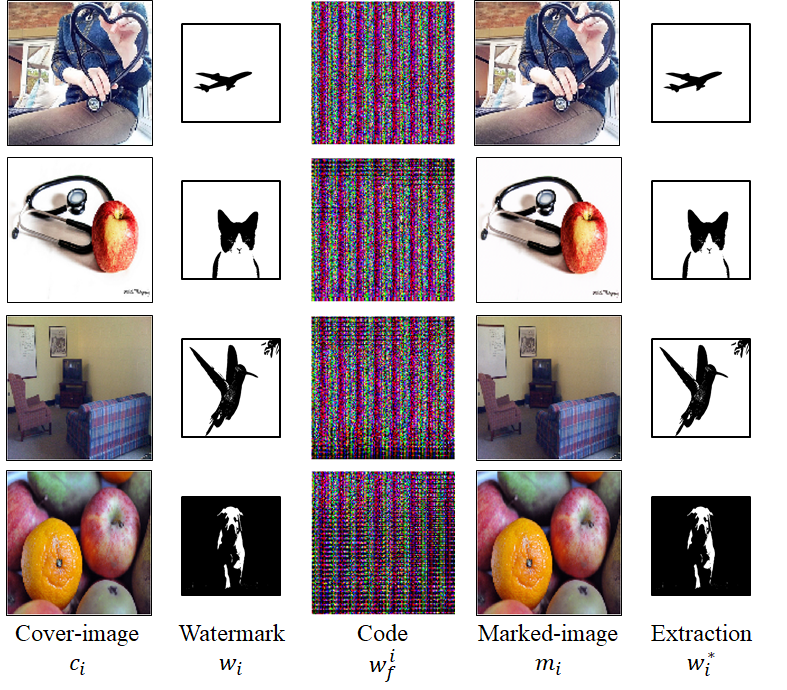}
    \caption{A few testing examples of the proposed scheme with various image content and color.}
    \vspace{-0.6cm}
    \label{fig:example}
\end{figure}

At the testing phase, the peak signal-to-noise ratio (\textit{PSNR}) and bit-error-rate (\textit{BER}) are respectively used to quantitatively evaluate the fidelity of the marked-images and the quality of the watermark extraction. The \textit{PSNR} is defined as:
\begin{equation}
   PSNR =10\log_{10} \left(\frac{max(c_i)^2}{MSE(c_i, m_i)} \right),
\end{equation}
where \textit{MSE} is the mean squared error. The \textit{BER} is computed as the percentage of error bits on the binarization of the watermark extraction $w_i^*$. In the testing, the \textit{BER} is zero, indicating that the original and the extracted watermarks are identical. The testing \textit{PSNR} is 39.72 dB, indicating a high fidelity of the marked-images, so that the hidden information cannot be noticed by human vision. A few testing examples with various image content and color are presented in Fig. \ref{fig:example}, where we can observe high fidelity and full extraction. The watermark codes $w_f^i$ do not directly reveal information about the watermark, which shows the perceivable randomness and decomposition learned by the proposed model. 

\subsection{The Proposed Scheme on Synthetic Images}\label{subsec:synthetic_images}
To further validate that the image watermarking task is properly generalized, the proposed scheme is applied for synthetic RGB cover-images and watermarks. The results of the blank cover-images and the random bits are presented below.

Fig. \ref{fig:rgb_example} illustrates the scenario of embedding binary watermarks into synthetic blank cover-images of black, red, white, green, and blue colors, respectively. Although the blank cover-images are not included in the training, the proposed scheme provides promising results. Applying blank cover-images is known to be extremely difficult in conventional watermarking methods due to the lack of psycho-visual information. However, unlike traditional methods that assign some unnoticeable portions of visual components as the watermark, the proposed deep learning model learned to apply the correlation between the features of space $W_f$ and of $M$ to indicate the watermark.

\begin{figure}[h!]
    \centering
    \includegraphics[width=0.9\linewidth]{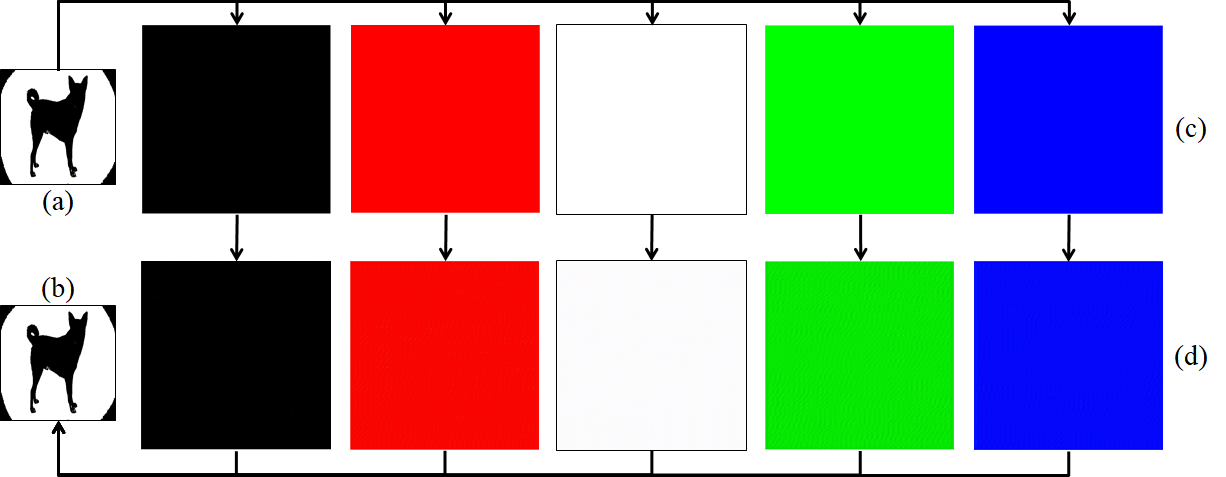}
    \caption{Embedding watermarks into blank cover-image examples. (a) and (b): the embedded and extracted watermarks; (c) and (d): the five blank cover- and marked-images.}
    \label{fig:rgb_example}
\end{figure}

\begin{figure}[h!]
    \centering
    \includegraphics[width=0.8\linewidth]{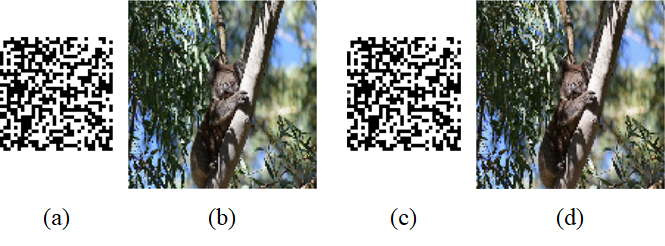}
    \vspace{-0.2cm}
    \caption{Embedding random bits as the watermark. (a) random bits; (b) a cover-image; (c) extraction; and (d) the marked-image.}
    \label{fig:randombits_example}
\end{figure}

Fig.~\ref{fig:randombits_example} shows an example of embedding a randomly generated binary image into a natural cover-image. To test the application scenarios where the watermarks are encrypted to random bits (besides the displayed example), $10,000$ randomly generated bit sets are tested on $10,000$ cover-images from the testing dataset. The average \textit{BER} is $0.36\%$, which indicates that applying random binary bits as the watermark does not deteriorate the performance of our proposed solution.

\subsection{The Robustness of the Proposed Scheme}\label{subsec:robust}

\begin{figure}[h!]
    \centering
    \includegraphics[width=0.75\linewidth]{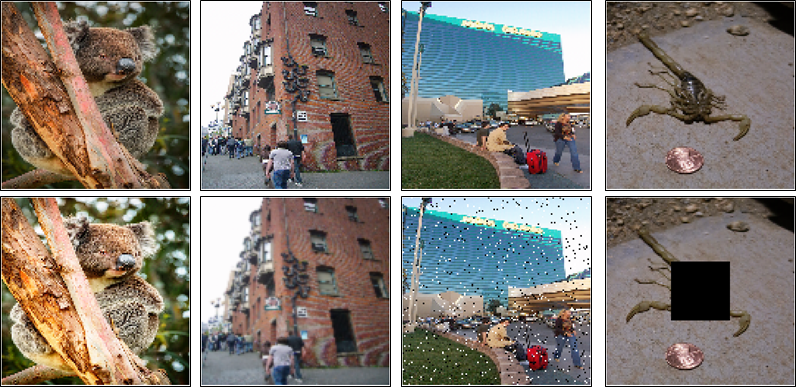}
    \caption{A few visual examples of the marked-images (top row) and their distortions (bottom row). The methods are implemented from left to right: Histogram Equalization, Gaussian Blur, Salt \& Pepper Noise, and Cropping, respectively.}
    \vspace{-0.3cm}
    \label{fig:robust_example}
\end{figure}

The robustness of the proposed scheme against different distortions on the marked-image is evaluated by analyzing the distortion tolerance range. Fig. \ref{fig:robust_example} illustrates a few visual examples of the marked-images and their distortions.

\begin{figure}[h!]
    \centering
    \includegraphics[width=0.8\linewidth]{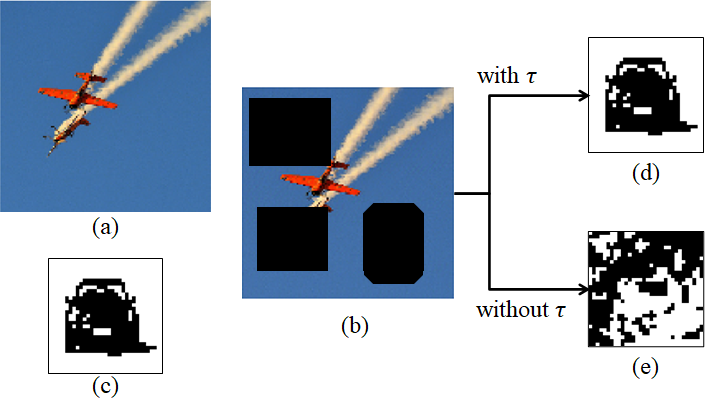}
    \vspace{-0.3cm}
    \caption{An example of watermark extraction under large percentage cropping. (a) an example marked-image, (b) cropped (a), (c) original watermark, (d) extraction with $\tau$, and (e) extraction without $\tau$.}
    \vspace{-0.2cm}
    \label{fig:example_cropping_ablation}
\end{figure}

Due to the over-complete design and the invariance layer $\tau_{\theta_5}$, the proposed schemes can tolerate distortions at a very high percentage (see Fig. \ref{fig:example_cropping_ablation} (b) and (d) for an example of large cropping).

\begin{figure}[h!]
    \centering
    \includegraphics[width=0.95\linewidth]{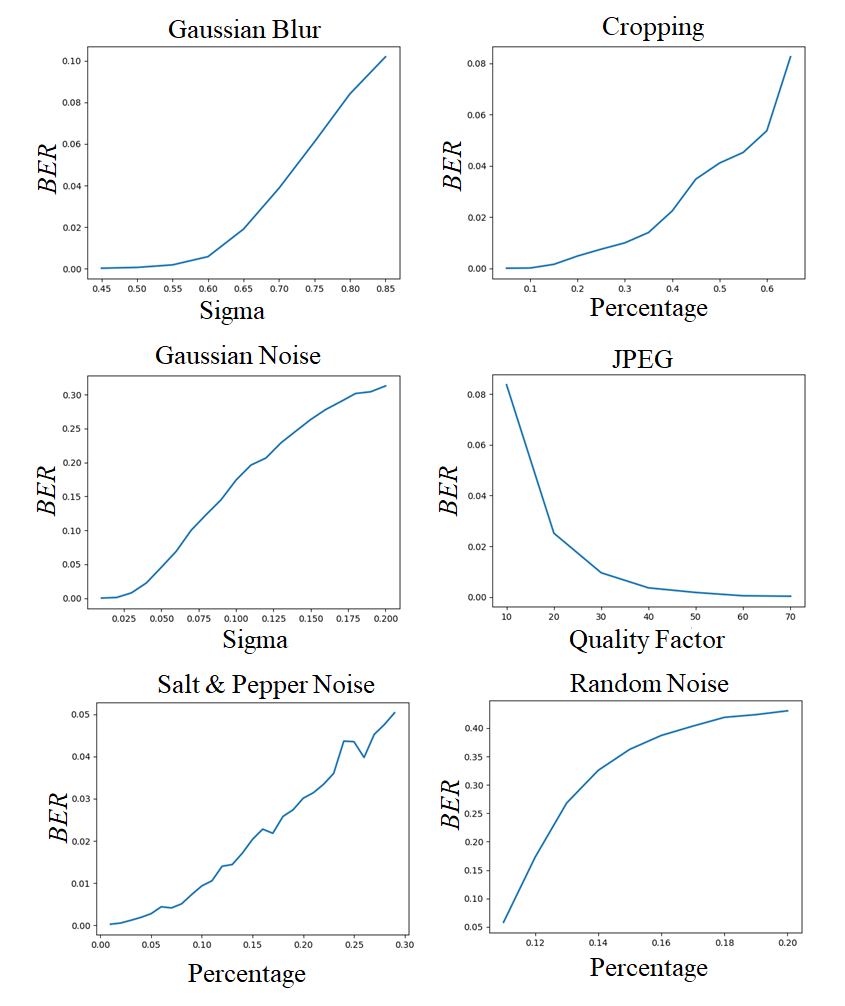}
    \vspace{-0.3cm}
    \caption{Distortions with swept-over parameters versus average BER.}
    \vspace{-0.3cm}
    \label{fig:robust_curve}
\end{figure}

To demonstrate the importance of our core robustness provider, two control experiments extracting watermarks from distorted marked-images with and without $\tau_{\theta_5}$ are conducted. Without $\tau_{\theta_5}$, the proposed scheme cannot extract the correct watermark (one example is shown in Fig.\ref{fig:example_cropping_ablation}), and the extraction from 10,000 testing attacked marked-images yields an average \textit{BER} as high as 42.46\%, which illustrates the significance of $\tau_{\theta_5}$ if we compare to the results presented in Fig.\ref{fig:robust_curve}.

\begin{figure}[h!]
    \centering
    \includegraphics[width=0.6\linewidth]{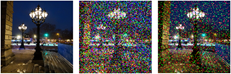}
    \caption{An example of extreme distortions. (left): the marked-image; (middle): after Gaussian additive noise with variance 0.2; and (right): after 20\% random noise.}
    \vspace{-0.3cm}
    \label{fig:extreme_fail}
\end{figure}

With $\tau_{\theta_5}$, distortions with swept-over parameters that control the attack strength are applied on the marked-images produced from the testing dataset. The watermark extraction \textit{BER} caused by each distortion under each parameter is averaged over the testing dataset. 
The distortions with swept-over parameters versus the average \textit{BER} are plotted in Fig. \ref{fig:robust_curve}. Since focusing on image-processing attacks, the responses of the proposed scheme against some challenging image-processing attacks are discussed. The proposed scheme shows high tolerance range on these challenges, especially for cropping, salt-and-pepper noise, and JPEG compression. 
For example, the extracted watermarks have low average \textit{BER}s as 7.8\%, 11.6\%, and 12.3\% under severe distortions including a cropping discarding 65\% of the marked image, a JPEG compression with a low quality factor 10, and a 90\% salt-and-pepper noise.
The attacks that randomly fluctuate the pixel values through image channels show a higher \textit{BER}, including Gaussian additive noise and random noise that sets a random pixel to a random value. 
These extreme attacks can easily destroy most of the contents on the marked-image (see few examples in Fig.~\ref{fig:extreme_fail}). Still, the proposed system achieves good performances when the marked-image contents are decently preserved, such as 14\% \textit{BER} on a 11\% random noise.

As discussed in Table~\ref{Table: analytic_comparison}, Mun’s scheme~\cite{mun2019finding} has the closest purpose with ours because it also achieves blindness and robustness simultaneously. Hence, we further compare our scheme with Mun's scheme for analysis. The comparison is performed on the same cover-image sets and the same watermark images as reported in Mun's scheme. To analyze the robustness of the proposed scheme, the extraction \textit{BER}s under common image-processing attacks are shown in Table~\ref{Table: quantitative_comparison}. The proposed scheme shows the advantages by covering more distortion categories in image-processing attacks and obtaining a lower \textit{BER} under the same distortion parameters. Although Mun's method can tolerate geometric distortions while the proposed scheme cannot, Mun's method requires the presence of the distortions in the training phase for robustness. In the real world, there is no way to predict and enumerate all kinds of attacks. 

\begin{table*}[h!]
\centering
\caption{A quantitative comparison between the proposed scheme and Mun's scheme.}
\label{Table: quantitative_comparison}
\begin{tabular}{lccccccc}
\hline \hline
\multirow{2}{*}{\textbf{Method}} & \multicolumn{4}{c}{\textbf{BER (\%) under the distortions}}                                                                                                                                                  & \multicolumn{1}{l}{}                                           & \multirow{2}{*}{\textbf{\begin{tabular}[c]{@{}c@{}}PSNR \\ (dB)\end{tabular}}} & \multirow{2}{*}{\textbf{\begin{tabular}[c]{@{}c@{}}Capacity\\ (bits)\end{tabular}}} \\ \cline{2-6}
                                 & \textbf{HE} & \textbf{\begin{tabular}[c]{@{}c@{}}JPEG \\ 10\end{tabular}} & \textbf{\begin{tabular}[c]{@{}c@{}}Cropping \\ 20\%\end{tabular}} & \textbf{\begin{tabular}[c]{@{}c@{}}S\&P \\ 5\%\end{tabular}} & \textbf{\begin{tabular}[c]{@{}c@{}}G. F. \\ 10\%\end{tabular}} &                                                                                &                                                                                     \\ \hline
Mun's                            & $N/A$          & $N/A$                                                         & 6.61                                                              & 7.98                                                         & 4.81                                                           & 38.01                                                                          & 256                                                                           \\
Ours                             & 0.43        & 8.16                                                        & 0                                                                 & 0.97                                                         & 0                                                              & 39.93                                                                          & 1,024                                                                               \\ \hline \hline
\end{tabular}
\vspace{0.5em}

\begin{tablenotes}
    \item Note: $N/A$ denotes "Not Applicable" that the robustness of an attack was not covered in~\cite{mun2019finding}; $HE$ denotes histogram equalization; JPEG 10 denotes a JPEG compression with quality factor 10; $S\&P$ denotes the salt-and-pepper noise; \textit{G. F.} denotes Gaussian filtering.
\end{tablenotes}
\end{table*}
\vspace{-0.5cm}

\subsection{A Case Study: Feasibility Test on Watermark Extraction from Camera Resamples}\label{subsec:feasibility_test}

\begin{figure}[h!]
    \centering
    \vspace{-0.3cm}
    \includegraphics[width=1.0\linewidth]{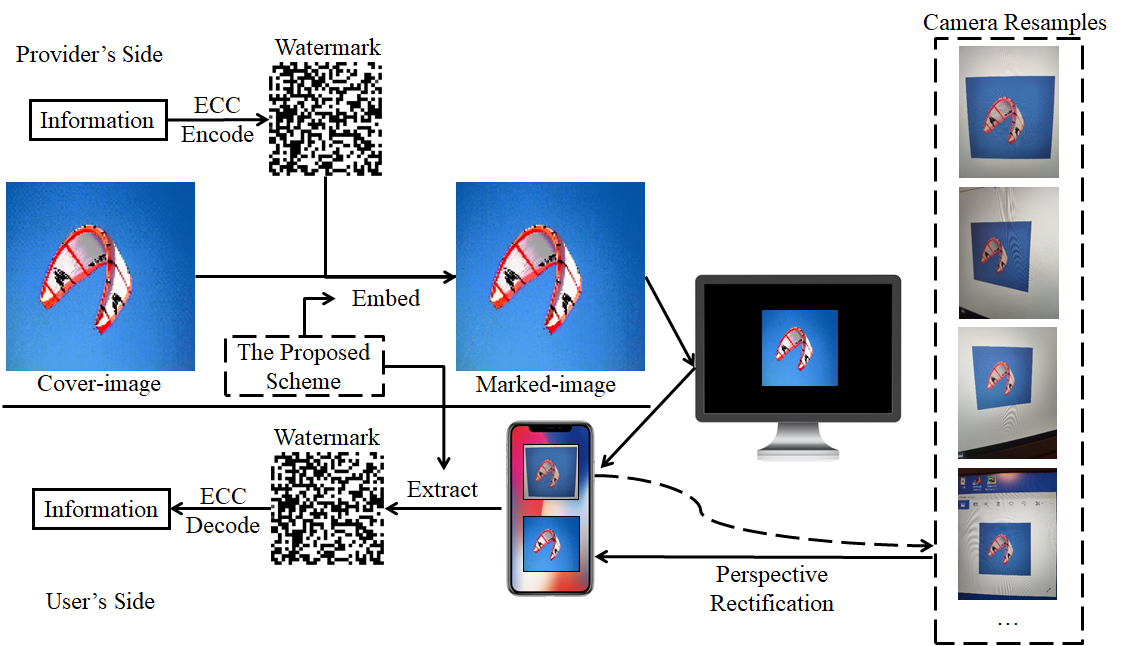}
    \vspace{-0.3cm}
    \caption{The phone camera testing scenario.}
    \vspace{-0.3cm}
    \label{fig:testing_scenario}
\end{figure}

Currently, image watermarking applications are much different than they used to be when they were first developed. Early scenarios focus on copyright detection, while more recent real-world communication requirements introduce a challenging use-case: the watermark extraction from phone camera resamples of marked-images~\cite{digimark}. The challenges in this use-case arise because watermark extraction needs to handle multiple combinations of the distortions, such as optical tilt, quality degradation, compression, lens distortions, and lighting variation. Most existing approaches~\cite{pramila2018increasing,kim2006image,yamada2013method,delgado2013digital} focus on the resamples of printed marked-images, not on phone resamples of a computer monitor screen. This use-case typically involves additional distortions, such as the Moir\'e pattern (\textit{i.e.,} the RGB ripple), the refresh rate of the screen, and the spatial resolution of a monitor (some examples are mentioned in Fig.~\ref{fig:testing_scenario}). We applied the proposed scheme as a major component in such scenarios since it is designed to reject all irrelevant noise instead of focusing on certain types of attacks. The outline of our scenario is shown in Fig.~\ref{fig:testing_scenario}.

An information provider prepares the information by encoding through an Error Correction Coding (ECC) technique. 
Although trained as an entire neural network, the proposed scheme is separated into the embedding components ($\mu_{\theta_1}$ and $\sigma_{\theta_2}$) and the extracting components ($\tau_{\theta_5}$, $\varphi_{\theta_3}$ and $\gamma_{\theta_4}$). The marked-image can be obtained by embedding the encoded watermark into the cover-image using the trained embedding components. The marked-image that looks identical to the cover-image is distributed online and displayed on the user's screen. A user scans the marked-image with a phone to extract the hidden watermark through the extracting components. 

The distortions occurred in this test can be divided into two categories: perspective and image-processing distortions. The major function of the proposed scheme in this scenario is to overcome the pixel-level modifications coming from image-processing distortions like compression, interpolation errors and the Moir\'e pattern. To concentrate the test on the proposed scheme, we simplified the solution of the perspective distortions, although this can be an entire challenging research track~\cite{zhao2018automatic, jaderberg2015spatial}.

\begin{figure}[h!]
    \centering
    \vspace{-0.3cm}
    \includegraphics[width=0.8\linewidth]{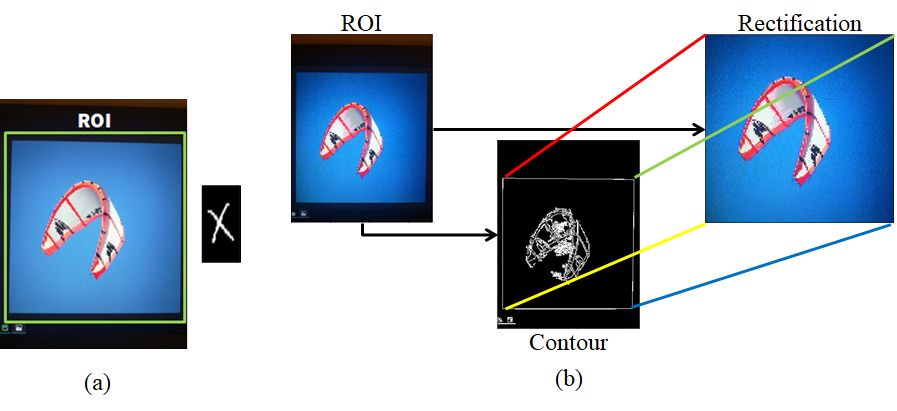}
    \vspace{-0.2cm}
    \caption{The prototype. (a) The GUI and sample information; and (b) the simple rectification.}
    \vspace{-0.4cm}
    \label{fig:prototype}
\end{figure}

With this setup, we develop a prototype for a user study. Fig.~\ref{fig:prototype} (a) illustrates the Graphical User Interface (GUI) and a 32 $\times$ 16 sample information. Classic Reed Solomon ($RS$) code~\cite{reed1960polynomial} is adopted as the ECC to protect the information. $RS(32, 16)$ is applied to protect each row of the $32 \times 16$ information so that the encoded information will be a $32 \times 32$ watermark satisfying the fixed watermarking capacity of the proposed scheme. In the watermark, each row is a codeword with data of length 16 and a parity of length 16, and hence can correct up to an error of length 8. Therefore, in this watermark of length 1,024 ($32 \times 32$), up to 256 errors can be corrected if there are no more than 8 errors in each row~\cite{reed1960polynomial}. Applying half of the bits as the parity, the watermarking payload is 512 bits. For the perspective distortion, four corners of the largest contour inside the Region Of Interest (ROI) are used to map the contoured content to a bird's-eye view (Fig.~\ref{fig:prototype} (b)), and the watermark is extracted from the bird's-eye view rectification.

In this case study, five volunteers were invited to take a total of $25$ photos of some marked-images displayed on a $2,560 \times 1,440$ screen using the camera on their mobile phones.
The volunteers' phones were Google Pixel 3, Samsung Galaxy s9, iphone XR, Xiaomi 8, and iphone X. All the photos were taken under office light conditions. Volunteers were given two rules.
First, the entire image should be placed as large as possible inside the ROI. As a prototype for demonstration, this rule facilitates our segmentation that the largest contour inside the ROI is the marked-image, so that this application can focus on the test of the proposed system instead of some complicated segmentation algorithms. In addition, placing the image largely in the ROI helps with the capture of desired details and features for the watermark extraction. Second, the camera should be kept as stable as possible. Although the proposed system tolerates some blurring effects, it is not designed to extract watermarks in high-speed motion.

\begin{figure}[h!]
    \centering
    \includegraphics[width=1.0\linewidth]{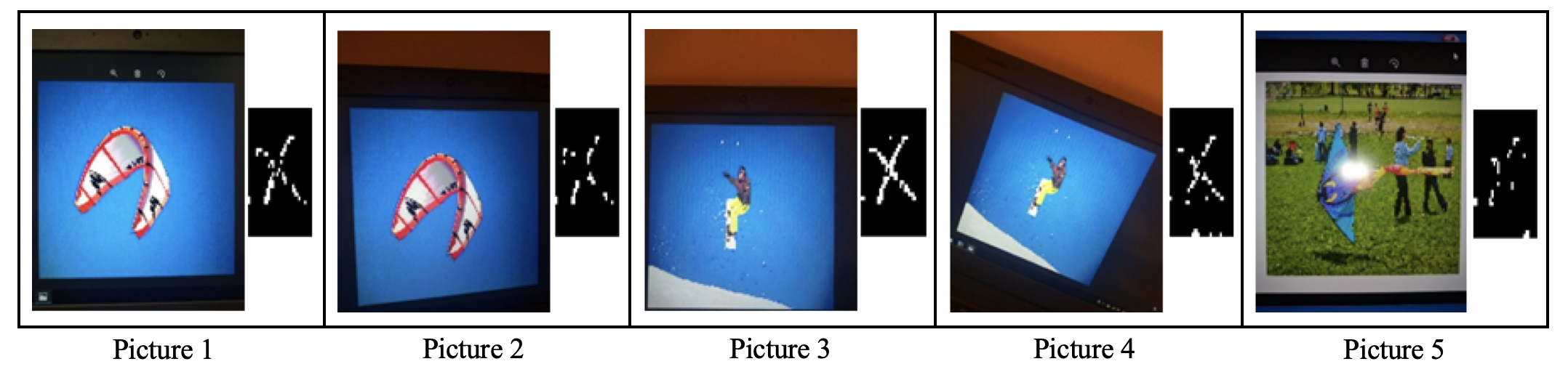}
    \caption{Five examples of the watermark extractions before ECC and their ROIs. The left-hand side of each picture is the ROI, and the right-hand side is the extracted watermark. The \textit{BER}s of the extractions from left to right are 3.71\%, 4.98\%, 1.07\%, 4.30\%, and 8.45\%, respectively.}
    \vspace{-0.4cm}
    \label{fig:camera}
\end{figure}

Fig.~\ref{fig:camera} presents five watermark extractions, their \textit{BERs}, and the corresponding ROIs. The closer up the photo is taken, the lower the error. Also, a lower error was observed with a greater parallel angle between the camera and the screen.
The flashlight brings more errors due to over- or under-exposure to some image areas. Use of the flashlight in this application is optional because the screen has the back-lights. The average \textit{BER} was 5.13\% for the $25$ images.

For a visual comparison, the displayed watermark extractions are the raw results before error correction. After executing $RS(32,16)$, all the watermark extractions in the testing cases can be restored to the original information without errors, as shown in Fig.~\ref{fig:prototype}. The proposed scheme can successfully extract the watermark within one second because it only applies the trained weights on the marked-image rectification.

%% file: Section/app.tex
In this section, we discuss different application scenarios, where the proposed image watermarking scheme can be applied to (i) authorized IoT device onboarding in Section~\ref{subsec:onboarding}; (ii) creation of private communication channels in Section~\ref{subsec:private_communication}; and (iii) authorized access to privileged content and services in Section~\ref{subsec:privileged_content_services}. We assume that a watermark is created based on credentials or secrets provided by users (e.g., passwords, cryptographic keys, or fingerprint scans). Our scenarios can also utilize the watermark extraction technique from camera resamples that we presented in Section~\ref{subsec:feasibility_test}.

\subsection {Authorized IoT Device Onboarding}
\label{subsec:onboarding}

IoT devices need to be "onboarded" once are bought by users in terms of being associated with a home controller or some cloud service so that users can control them~\cite{mastorakis2020icedge, mastorakis2019towards}.

Currently, the widely-used methods to onboard IoT devices are mostly out-of-band and include the use of QR codes physically printed on devices, pin codes, and serial numbers~\cite{latvala2020evaluation}. For example, once a user buys a smart IoT camera, he/she scans a QR code printed on the camera (or the packaging) with his/her mobile phone and through a mobile application, he/she connects the camera with a cloud service usually offered by its manufacturer. In this way, the user is able to watch the video stream capture from the camera.

Existing onboarding methods do not protect against unauthorized access. For example, attackers that have physical access to an IoT device can tamper it (e.g., install malware on the device) before the device is onboarded by its owner. The proposed image watermarking mechanism can be utilized to enable the onboarding of IoT devices only by the device owner. For example, user credentials can be embedded to a QR code, which will be physically printed on a device. Once a user receives his/her IoT device (\textit{i.e.}, an IoT camera with a QR that has the user credential embedded), he/she takes a picture of the QR code with his/her mobile phone. To onboard the device, the user sends the taken QR picture along with his/her credentials to a server that runs the extraction via a deep neural network. The deep neural network verifies that the user indeed possesses the credentials (watermark) embedded into the QR code and authorizes the user to onboard the IoT device. 

\subsection {Creation of Private Communication Channels}
\label{subsec:private_communication}

The proposed image watermarking scheme can be used for the creation of private chatrooms and other communication channels. For instance, a chatroom organizer can collect the credentials of individuals that he/she would like to communicate with and create a QR code with this set of credentials embedded to the code. The created QR code can be uploaded on the Internet. Once an individual that has been included in the communication group scans the QR code with his/her mobile phone and provides his/her credential to the deep neural network, the network verifies that this individual indeed possesses credentials embedded into the QR code and authorizes the user to join the chatroom. Unauthorized Internet users (i.e., users that do not have their credentials embedded into the QR code watermark) might try to join the chatroom, but they will not be able to do so; even if such users take a photo of the QR code, they do not possess credentials that are embedded into this code; thus, the deep neural network will reject their requests to join the chatroom. The QR code with the embedded watermark will be publicly available and can be accessed by all Internet users, however, only the authorized users will be able to join the chatroom and communicate with each other.

\subsection {Authorized Access to Privileged Content and Services}
\label{subsec:privileged_content_services}

The proposed image watermarking scheme can be also utilized for a broader scope of applications, where access to privileged content and/or services is desirable. The content producer or service provider will create cover-images that have the credentials of authorized users embedded. As a result, image watermarking can be used as an access control mechanism, where access to certain pieces of (privileged) content and/or services are restricted only to authorized users; \textit{i.e.}, users that have their credentials (watermark) embedded into the marked image.
Similarly, when authorized users send the marked image and the embedded credentials to the deep neural network, the network will be able to extract the watermark only if the user possesses the proper credentials. Only users that possess the credentials (watermark) embedded into the marked image will be allowed to access the privileged content. 

%% file: Section/conclusion.tex
This paper introduces an automated and robust image watermarking scheme using deep convolutional neural networks. The proposed blind image watermarking scheme exploits the fitting ability of deep neural networks to generalize image watermarking algorithms, shows an architecture that trains in an unsupervised manner for watermarking tasks, and achieves its robustness property without requiring prior knowledge of possible distortions on the marked-image. Experimentally, we have not only reported the promising performances for individual common attacks, but also have demonstrated that the proposed scheme has the ability and the potential to help combinative, cutting-edge, and challenging camera applications, which has confirmed the superiority of the proposed scheme. Our future work includes tackling geometric and perspective distortions by the deep neural networks inside the scheme, and refining the scheme architecture, objective and loss function by different methods like ablation studies.